\documentstyle[11pt,newpasp,twoside,epsf]{article}
\def\edcomment#1{\iffalse\marginpar{\raggedright\sl#1\/}\else\relax\fi}
\reversemarginpar

\newcommand{\be}{\begin{equation}}
\newcommand{\ee}{\end{equation}}
\newcommand{\bea}{\begin{eqnarray}}
\newcommand{\eea}{\end{eqnarray}}
\newcommand{\half}{{\textstyle \frac{1}{2}}}

\newcommand{\fat}[1]{{\bf #1}}
\newcommand{\doo}[2]{{\frac{\partial #1}{\partial #2}}}
\def\ZZ#1{$\scriptstyle #1$}

\begin{document}
\title{Regularization tools for binary interactions}
 \author{Sverre J. Aarseth}
\affil{Institute of Astronomy, University of Cambridge, Cambridge, UK}

\begin{abstract}
We first discuss two-body and chain regularization methods for
direct N-body simulations on HARP-2 and GRAPE-6.
The former is used for accurate integration of perturbed binaries and
hierarchies, whereas the latter deals with strong interactions involving
binaries.
Combined with a powerful stability criterion for hierarchical
systems, this versatile treatment provides an efficient way of
studying all dynamical processes in globular clusters containing
a realistic population of primordial binaries.
These algorithms are also ideal for modelling a variety of astrophysical
effects, such as averaging over Kozai cycles, tidal circularization,
spin-orbit coupling and stellar collisions, where well-defined
elements are used to describe near-singular solutions.
We also describe a new time-transformed leapfrog scheme which has been
developed to deal with black-hole binaries in galactic centres.
This formulation is valid for large mass ratios and dominant two-body
motions in a compact subsystem can be treated accurately in the
post-Newtonian approximation.
This method has been combined with the existing regularization algorithms
into a new simulation code {\ZZ {NBODY7}}.
Some preliminary results of a test problem illustrating possible
applications are presented.
\end{abstract}

\section{Introduction}

The presence of binaries in star cluster simulations poses many technical
questions which must be overcome in order to obtain satisfactory results.
In the following, we review the main algorithms for studying a variety of
interactions, ranging from physical collisions to stable hierarchies.
It is somewhat paradoxical that the degree of difficulty in dealing with
these two situations is such that the former is by far the easiest.
Given a significant binary fraction, it is necessary to treat close
encounters of three or more strongly interacting particles.
This is achieved by three-body and chain regularization.
In addition, stable configurations of three or more members form by various
processes.
These hierarchical structures range from triples to sextuplets.
Upon ascertaining stability, such systems are described by two-body
regularization.
Here the inner binary may be subject to induced eccentricity increase due
to the inclination effect, which can also be considered.
With all these tools, it is now possible to study the long-term evolution of
conventional star clusters, taking account of all relevant astrophysical
processes.
However, the modelling of black-hole binaries reveal some inefficiencies
during the advanced stages of evolution.
A new method is presented for dealing with this challenging problem.

\section{Primordial Binaries}

\begin{figure}
\plotfiddle{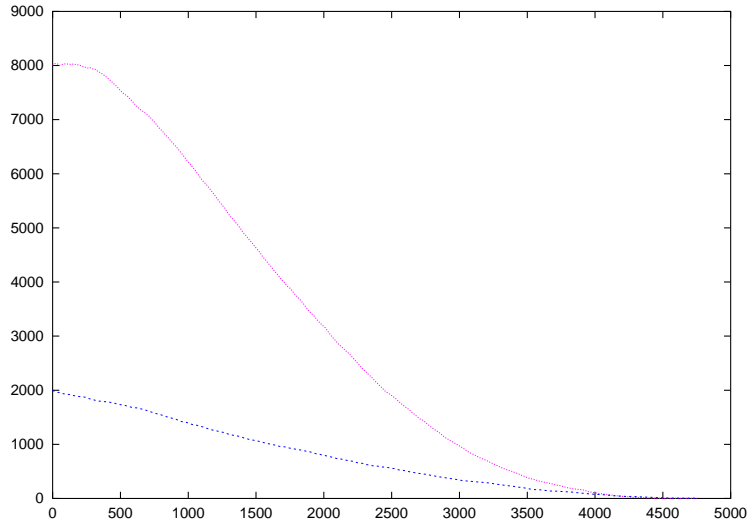}{6.5cm}{270}{40}{40}{-160}{220}
\caption{Number of single stars and binaries ($N$-body units).}
\end{figure}

Realistic star cluster simulations usually include a significant population
of primordial binaries.
We concentrate on hard binaries and choose a flat distribution of semi-major
axes (Kroupa 1998).
These binaries are integrated by the classical Kustaanheimo-Stiefel
(1965, hereafter KS) regularization method.
After trying several alternatives, the Stumpff Hermite formulation appears
to be superior (Mikkola \& Aarseth 1998) and yields high accuracy at small
perturbations.
The main idea here is that correction terms are included for the neglected
higher orders.
In addition, the so-called slow-down procedure which exploits the adiabatic
invariance (Mikkola \& Aarseth 1996) allows one KS orbit to represent several
physical periods by scaling the perturbing force and physical time
accordingly.
At even smaller perturbations, the two-body orbit is assumed to be
conserved over one or more periods, depending on the nearest neighbours.
This approximation leads to a considerable speeding-up of the integration
since the binary can now be treated as a single particle.

Use of regularized two-body descriptions facilitates implementation of
astrophysical processes connected with finite-size effects (Aarseth 1996).
Among these are tidal capture and circularization (Mardling \& Aarseth 2001),
Roche-lobe overflow, common-envelope evolution, magnetic or relativistic
braking and physical collisions (Tout et al. 1997).
In addition, slow expansion due to stellar mass loss or disruption following
a neutron star velocity kick is also treated.
An energy-conserving scheme is achieved by including a variety of
correction procedures.
Further details about the different processes have been reviewed
elsewhere (Aarseth 1999a).

KS regularization has been implemented in several direct summation codes.
Here we mention {\ZZ {NBODY4}} for use on the special-purpose HARP-2 or
GRAPE-6 computers and {\ZZ {NBODY6}} which is suitable for laptops and
workstations (Aarseth 1999b).
The former is based on full $N$ force summation, whereas the latter employs
the Ahmad-Cohen (1973) neighbour scheme for increased efficiency.
Both these recent versions use Hermite integration for the single particles
(Makino 1991).
With the special-purpose machines, the force on a binary centre of mass is
also obtained by direct summation, followed by differential correction on
the host due to the perturbers.
Typical models of open clusters studied on HARP-2 consist of $N_s = 8000$
single stars and $N_b = 2000$ hard binaries, with total life-times exceeding
5~Gyr (cf.\ Hurley et al. 2001).
Figure 1 shows the population of bound single stars and binaries as a
function of time in a standard open cluster model.
Near half-life at $t = 1690$ (or $\simeq 2.2$~Gyr), $N_s = 4035$ and
$N_b = 957$; hence the binary population is not preferentially depleted.

\section{Hierarchical Systems}

Observations of the solar neighbourhood reveal a significant fraction of
triples as well as systems of higher order.
Likewise, analysis of star cluster models also show that such configurations
are not rare (Aarseth \& Mardling 2001).
The main channel for formation of triples is by binary-binary collisions
and may be represented symbolically by
\be
B + B \Rightarrow [B,S] + S \,.
\ee
This important process was already identified during scattering experiments
where one single star is ejected (Mikkola 1983).

Depending on the orbital characteristics, an isolated triple may be stable
and the life-time may still be considerable in the presence of perturbers.
We test stability by a semi-analytical criterion for the outer pericentre
given by (Mardling \& Aarseth 1999)
\be
{R_p^{\rm crit}} ~=~ C\left[(1+q_{\rm out})\frac{(1+e_{\rm out})}
{(1-e_{\rm out})^{1/2}}\right]^{2/5}a_{\rm in} \,,
\ee
Here $C \simeq 2.8$ is a constant, $q_{\rm out}$ and $e_{\rm out}$ is the
outer mass ratio and eccentricity, respectively, and $a_{\rm in}$ the inner
semi-major axis.
Consequently, a triple system is said to be stable, provided
$a_{\rm out} (1 - e_{\rm out}) > R_p^{\rm crit}$.
This criterion has been checked extensively and appears to be quite robust.
It has been generalized to the case of an outer binary and an inclination
correction factor has also been included to represent the enhanced stability
of retrograde orbits.
The formation mechanism of quadruple systems is still under investigation
but appears to be connected with multiple encounters.
Here the early stage is characterized by a relatively soft binding energy,
whereas many triples are already hard at the outset.

Once accepted as stable, a hierarchy is described by two-body motion and
studied by KS regularization until the criterion is violated or the external
perturbation dictates termination.
In spite of the small formation rate, the population of hierarchical
systems becomes significant after the early epoch, with typically a dozen
members (Aarseth \& Mardling 2001).
Temporary high-order systems of the type
$[[B,S],S], \,[[B,S],B], \,[[B,B],B]$ occur and it is not uncommon even for
quadruples to escape.
Figure 2 illustrates the number of stable hierarchies as a function of time
in the model used for Fig.~1.

\begin{figure}
\plotfiddle{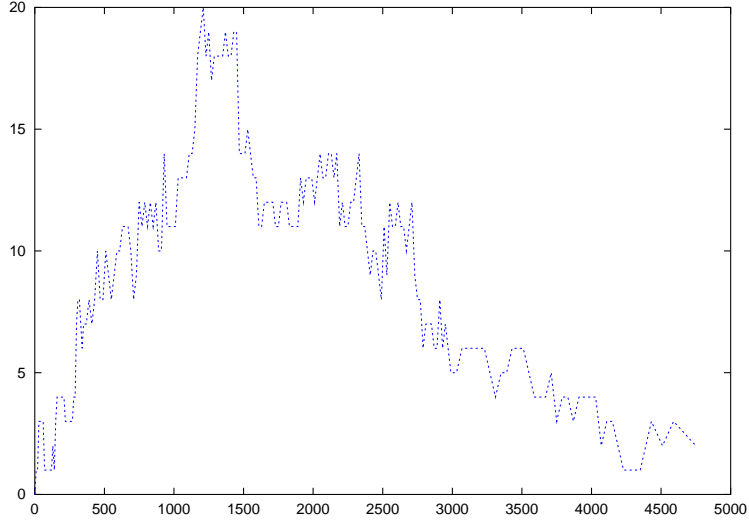}{6.5cm}{270}{40}{40}{-160}{220}
\caption{Stable hierarchical systems.}
\end{figure}

An attempt has been made to model the Kozai (1962) cycles of the inner
binary for favourable values of the inclination (Mardling \& Aarseth 2001).
Thus provided the maximum eccentricity reaches a sufficiently high value,
tidal dissipation may be activated and initiate circularization with
consequent orbital shrinking.
We include the effect of stellar rotation (Hut 1981) which tends to advance
the onset of Roche-lobe mass transfer by the synchronization process.
The averaging treatment integrates the eccentricity and angular momentum
vectors together with contributions from the outer body in the quadrupole
approximation (Heggie 1996, Aarseth \& Mardling 2001).
Here we also include apsidal motion and relativistic precession if relevant.
The new two-body elements are converted to relative coordinates and
velocities, followed  by a transformation to KS variables for the inner
binary which remains dynamically inert.
In spite of various de-tuning effects, some examples of tidal circularization
are in fact seen.
Note that in order for this to occur, the eccentricity needs to reach quite
high values; i.e. $e_{\rm max} > 0.999$.

\section{Strong Interactions}

The presence of hard binaries invariably leads to close encounters involving
large energy changes.
Compact subsystems with 3-5 members are treated by chain regularization
which also includes a slow-down procedure for super-hard binaries
(Mikkola \& Aarseth 1993, 1996).
In addition, unperturbed three-body regularization (Aarseth \& Zare 1974)
is available during an ongoing chain interaction.
Although there are only $\sim 10^3$ strong interactions during a typical
cluster life-time, many such events are associated with energetic ejections
of single stars as well as binaries.
Thus escape velocities $\sim 100~{\rm km~s}^{-1}$ may be seen in cluster
models with rms velocity $\simeq 2~{\rm km~s}^{-1}$.

Close binary-binary encounters often lead to the formation of triples, many
of which are stable.
In the alternative case of ejection, the remaining binary tends to acquire
high eccentricity with possible collision.
According to the results, the latter outcome occurs more frequently than
hyperbolic two-body collisions, whereas tidal capture is seen on rarer
occasions.
The initiation of normal or chaotic tidal interactions is also connected
with multiple encounters involving one or two binaries.
Again all the astrophysical processes are treated using well-defined
quantities.
In the case of a physical collision, this requires careful iteration of
the numerical solution which employs the Bulirsch-Stoer (1966) integrator.
Here the osculating semi-major axis is obtained after evaluating the
non-singular interactions, combined with the total subsystem energy.

Implementation of chain regularization in {\ZZ {NBODY4}} and {\ZZ {NBODY6}}
requires considerable software development for decision-making.
Here we need to combine the integration of regularized solutions based on
non-linear time-steps with the Hermite scheme for commensurate steps.
After nearly ten years experience, these procedures are now working quite
well and are able to deal with most configurations.
It is particularly important to identify stable hierarchies in order to
enable a more efficient description.
Since external perturbations are included, a long-lived system may cross the
stability boundary slowly; hence frequent checks are needed.

The few-body regularization methods described above are also ideally suited
to the study of small systems ($N \le 10$) as well as for classical
scattering experiments.
In particular, more simulations are needed to explore the parameter space
for binary-binary scattering.
However, depending on the termination conditions, it may be necessary to
combine this approach with the variation of elements method or assume
unperturbed two-body motion for weakly bound ejections (Mikkola 1983).

\section{Super-Massive Binary Method}

The regularization schemes discussed above work quite well for most stellar
systems.
However, the case of very large mass ratios requires special consideration
in order to avoid the loss of efficiency and accuracy.
The problem of black-hole binaries in galactic nuclei is topical and has
received much attention.
Direct integration on GRAPE-4 with $N = 16,384$ and a component mass ratio
of 256 yielded some interesting results (Makino \& Ebisuzaki 1996).
Here the dominant interaction was calculated as point-masses, using full
double precision on the host computer.
The direct approach leads to difficulties for small pericentre distances
and in any case this investigation did not describe extreme evolution.
Attempts to use KS regularization have only been partially successful
(Quinlan \& Hernquist 1997, Milosavljevi\'c \& Merritt 2001).
One difficulty here is trapping of bound stellar orbits with respect to one
of the massive binary components without resulting in coalescence.
On general grounds, extension to chain regularization is also unlikely to be
satisfactory because the total subsystem energy which appears explicitly in
the equations of motion is dominated by the binary components.
In view of the small period and large binding energy of such a binary, some
other kind of regularization method is therefore desirable.

In the following we describe a time-transformed leapfrog scheme which offers
some practical advantages in dealing with large mass ratios
(Mikkola \& Aarseth 2001).
Consider first the standard leapfrog equations
\bea
\fat r_{\half} &=& \fat r_0+\frac{h}{2} \fat v_0 \nonumber \\
\fat v_1 &=& \fat v_0+h\:  \fat F(\fat r_{\half}) \nonumber \\
\fat r_1 &=& \fat r_{\half}+\frac{h}{2} \fat v_1 \,,
\eea
where $h$ is the time-step and ${\bf F}$ denotes the acceleration at
$t = \half h$.
We adopt a time transformation $ds = \Omega({\bf r}) dt$, with $\Omega$ an
arbitrary function and introduce a new auxiliary quantity $W = \Omega$.
The new idea here is to evaluate $W$ by
\be
\dot W = \fat v\cdot \doo{\Omega}{\fat r} \,.
\ee
This allows us to solve two sets of equations; namely (i)
${\bf r}' = {\bf v}/W, \, t' = 1/W$, and (ii)
${\bf v}' = {\bf F}/\Omega, \, W' = {\dot W} /\Omega$.
Consequently, we write
\bea
{\bf r} &=& {\bf r}_0 + s {\bf v}/W \nonumber \\
 t &=& t_0 + s/W  \,,
\eea
and
\bea
 {\bf v} &=& {\bf v}_0 + \frac {s \, {\bf F}} {\Omega} \nonumber \\
 W &=& W_0 + s \,\frac {{\bf v + v}_0} {2 \Omega} \cdot \doo{\Omega}{\fat r}
 \,.
\eea
Here the solution may be combined into a form of leapfrog, taking
$s = \half h$ in (5); then $s = h$ in (6) and finally again
$s = \half h$ in (5).

The numerical solution is obtained by the Bulirsch-Stoer (1966) rational
function extrapolation method.
Thus, using the above leapfrog algorithm, several integrations are performed
with gradually decreasing substeps $h$ and the results are extrapolated to
zero steplength.
This formulation has been generalized to include external perturbations of
conservative type.
An iterative procedure is necessary for treating the velocity-dependent
relativistic corrections and the post-Newtonian approximation (PPN2.5) has
been implemented.
In either case, the energy of the subsystem is no longer constant but its
change can be determined by straightforward integration.

Simple tests indicate that the time-transformed leapfrog (TTL) method is
very promising.
Thus significantly higher accuracy with about half the number of function
evaluations was achieved when including the time transformation.
This test employed the so-called Pythagorean three-body problem but it is
desirable to examine other initial conditions.
Moreover, reliable solutions for coalescence by gravitational radiation have
been obtained involving one or both of the massive binary components.
Since the Bulirsch-Stoer integrator is rather expensive when including many
interactions, the new method is intended for treating a compact subsystem
containing a massive binary and significant perturbers.
Finally, we note that integration of the time transforming function $W$
ensures a more well-behaved solution than direct evaluation.

\section{NBODY7: A New Simulation Code}

\begin{figure}
\plotfiddle{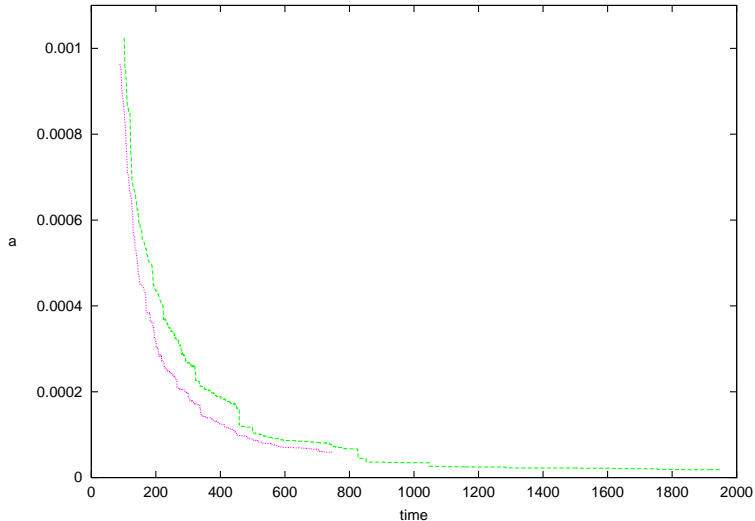}{6.5cm}{270}{40}{40}{-160}{220}
\caption{Evolution of semi-major axes; $N = 5000$.}
\end{figure}

The TTL method has already been combined with the star cluster code
{\ZZ {NBODY6}}.
In view of the many different procedures, it will be known as {\ZZ {NBODY7}}.
Consequently, all the standard regularization methods are still included for
increased flexibility.
The special treatment of a super-massive binary may begin with an existing
small semi-major axis as provided by another calculation, possibly proceeded
by a stage of KS regularization (cf.\ Milosavljevi\'c \& Merritt 2001).

In the following, we consider the more general case of two single massive
members located at arbitrary positions with typical velocities.
Depending on the mass ratio and particle number, the two massive bodies tend
to lose kinetic energy by dynamical friction and spiral to the centre on the
mass segregation time-scale (Hemsendorf, Sigurdsson \& Spurzem 2001).
At some stage, the components will form a binary and continue to shrink by
ejecting other particles from the core until the conditions are suitable for
KS regularization.
Hence selection for treatment by the TTL method is postponed until the
semi-major axis, $a_{\rm bh}$, reaches the characteristic size for a hard
binary.
The reason for the delay is that the Bulirsch-Stoer integrator only requires
a few steps during each Kepler period and this would cause prediction
problems for neighbouring particles which tend to have somewhat smaller
time-steps by the Hermite method.

The presence of a central subsystem requires special decision-making in order
to maintain an appropriate membership.
This entails perturber selection as well as increasing the subsystem
membership in a similar way as in KS and chain procedures.
Likewise, ejected subsystem particles must be removed and re-initialized as
field particles.
However, use of a fixed boundary is too simplistic since some particles may
have their apocentre just outside the critical value, with nearly constant
semi-major axis.
Typically, field particles inside $50 \, a_{\rm bh}$ are selected as
perturbers, with each one having a small effect because of the mass ratio.
Moreover, particles inside $20 \, a_{\rm bh}$ are chosen as members of the
subsystem itself.
In order to prevent numerical problems, a softened potential may be adopted
for interactions between the latter, whereas all terms involving the binary
components are treated in the point-mass limit.

\begin{figure}
\plotfiddle{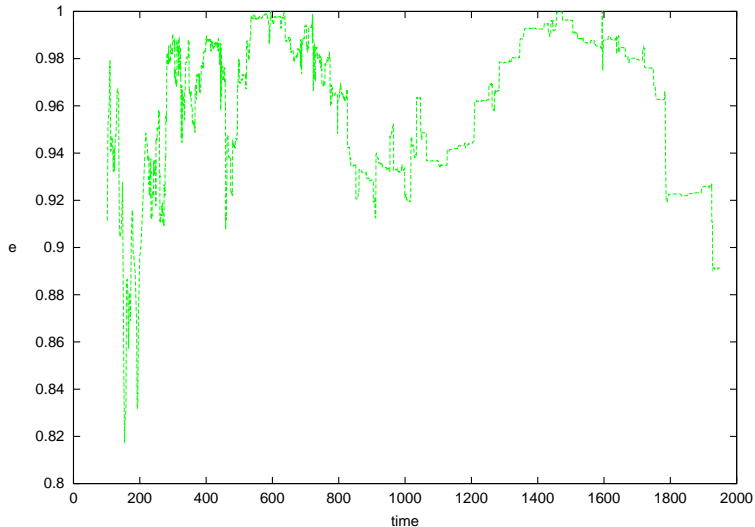}{6.5cm}{270}{40}{40}{-160}{220}
\caption{Eccentricity evolution.}
\end{figure}

In the case of an ultra-hard binary containing black-hole components, the
PPN2.5 post-Newtonian approximation is included for the two-body interaction.
The full expression is given by (Soffel 1989)
\be
{\bf F} = {\bf F}_0 + c^{-2}\,{\bf F}_2
+ c^{-4}\,{\bf F}_4 + c^{-5}\,{\bf F}_5 \,,
\ee
where ${\bf F}_0$ denotes the Newtonian force per unit mass and $c$ is the
speed of light.
Depending on circumstances, the other terms which are fairly time-consuming
may be omitted.
Alternatively, the relativistic radiation (${\bf F}_5$) may be included
without contributions from the precession terms.
There are two additional cases when the relativistic terms are considered.
First, this is done if the osculating pericentre of the black-hole binary
falls below a suitably small value related to the Schwarzschild radius.
We note that the question of possible high eccentricity for realistic
systems is still unresolved because of scaling uncertainties.

In the second critical case, the full GR treatment is activated if one of
the subsystem members has a small impact parameter with respect to either
component.
Accordingly, a temporary switch from the dominant members is made to the new
configuration, $m_i,\,m_k$, which would permit coalescence if relevant.
Here we have adopted a provisional collision condition given by
\be
r_{\rm coll} = 6 \, G \, (m_i + m_k) /c^2 \,.
\ee
Numerical tests have been performed with artificially small values of $c$,
demonstrating the correctness of the coalescence procedure.

A modest test calculation has been performed on a workstation in order to
illustrate some features of the method.
Initial conditions were chosen from a Plummer model containing $N = 5000$
equal-mass particles, with the two first being assigned intermediate masses
$N^{1/2} \bar m$ as a good compromise.
Figure 3 shows the evolution of the dominant binary during 2000 $N$-body
time units.
A similar second example was also studied for comparison, again without any
artificial scaling.
In the first example, the KS solution was initialized at
$a_{\rm bh} = 1.0 \times 10^{-3}$, followed by the TTL treatment from
$a_{\rm bh} = 3.8 \times 10^{-4}$ until $1.8 \times 10^{-5}$.
The slow rate of shrinkage seen during the later stages is partly due to
the depleted central density.

The behaviour of the eccentricity is displayed in Fig.~4.
Two epochs of extremely high eccentricity can be seen.
These maxima are associated with Kozai (1962) cycles (without GR) due
to field particles in stable orbits with inclination near $90^{\circ}$.
Depending on the density, such configurations will eventually be modified
by external perturbations.
Other trapped particles experience slingshot interactions and are ejected
with large velocities.
The corresponding recoil effect can be seen in Fig.~5 which illustrates
the time evolution of the square space velocity.
In particular, there are several spikes in addition to a number of
relatively large values.
Here the effective mass ratio is 140, with an initial mean square velocity
of 0.5.
Consequently, the resulting binary recoil leads to significant core
wandering which has been noted before.
As for possible loss-cone replenishments, we note that the central restoring
force may also be weak in larger systems, thereby enabling significant
excursions to take place.

\begin{figure}
\plotfiddle{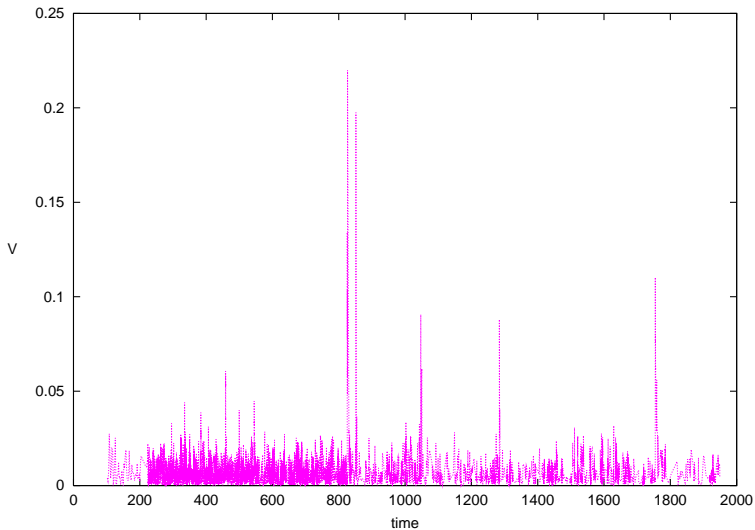}{6.5cm}{270}{40}{40}{-160}{220}
\caption{Square space velocity of massive binary.}
\end{figure}

\section{Conclusions}

The preceding sections have reviewed several regularization techniques which
have proved themselves in star cluster simulations (cf.\ Hurley et al.\ 2001).
Although some sophisticated decision-making is required for efficient use,
such formulations improve the numerical accuracy and also enable many
astrophysical processes to be studied by well-behaved variables.
In the second part, we described a new method for treating a super-massive
binary and its significant perturbers, with the addition of post-Newtonian
terms where relevant.
This formulation can be used to study compact subsystems in galactic nuclei.
So far, the TTL scheme has been implemented in a star cluster code for
workstations.
Given the existing procedures, very little effort is required to combine it
with the current {\ZZ {NBODY4}} code.
This timely development should allow much more realistic simulations on the
powerful GRAPE-6 special-purpose computer.

\end{document}